\documentstyle[epsf]{article}
\begin{document}

\begin{center}

{\Large {\bf 
               Populations of close binaries 
               in galaxies with recent  
               bursts of starformation

\bigskip
\bigskip

S.B. Popov, \\
}}

{{\large \bf (http://xray.sai.msu.su/\~ \, polar/)}}\\

{\Large {\bf

M.E. Prokhorov \& V.M. Lipunov \\

}}

\bigskip

{{\large \bf

Sternberg Astronomical Institute\\

Moscow State University\\
}}

\end{center}

\bigskip

{\Large

\centerline{{\bf Abstract}}
}

\bigskip

{\large

    This paper is a continuation and development of our previous
articles (Popov et al., 1997, 1998).
We use ``Scenario Machine'' (Lipunov et al., 1996b) -- the population synthesis
simulator (for single binary systems calculations 
the program is available in WWW:\\
{\bf http://xray.sai.msu.su/sciwork/scenario.html}
(Nazin et al., 1998) -- 
to calculate evolution of populations of several
types of X-ray sources during the first 20 Myrs after a starformation burst.

We examined the evolution  
of 12 types of X-ray sources in close binary systems (both with neutron
stars and with black holes) for different parameters of the IMF -- slopes: 
$\alpha=1$, $\alpha=1.35$ and $\alpha=2.35$ and
upper mass limits: 120 $M_{\odot}$, 60 $M_{\odot}$
and 40 $M_{\odot}$. Results, especially for sources with black holes, are
very sensitive to variations of the IMF, and it should be taken into account
when fitting parameters of starformation bursts.

Results are applied to several regions of recent starformation in
different galaxies: Tol 89, NGC 5253, NGC 3125, He 2-10, NGC 3049. 
Using known ages and total masses of starformation
bursts (Shaerer at al., 1998)
we calculate expected numbers of X-ray sources in close binaries for
different parameters of the IMF. Usially, X-ray transient sources consisting
of a neutron star and a main sequence star are most abundant, but for very
small ages of bursts (less than $\approx 4$ Myrs) 
sources with black holes can become more abundant. 
}

\newpage

\section{Introduction.}

 Theory of stellar evolution and one of the strongest tools 
of that theory -- population synthesis -- 
are now rapidly developing branches of astrophysics.
Very often  only the evolution of single stars is modelled,
but it is well known
that about 50\% of all stars are members of  binary systems,
and a lot of different astrophysical objects are products
of the evolution of binary stars. We argue, that often it is
necessary to take into account the evolution of close binaries
while using the population synthesis in order to avoid serious  errors. 

 Initially this work was 
stimulated by the article Contini et al. (1995),
where the authors suggested an unusial form of the initial mass function
(IMF) for the explanation of the observed properties
of the  galaxy Mrk 712 . They suggested the ``flat'' IMF with the exponent  
$\alpha=1$ instead of the Salpeter's value  $\alpha=2.35$.
Contini et al. (1995) didn't take into account binary systems, so
no words about the influence of such IMF  
on the populations of close binary stars could be said.
Later Shaerer (1996) showed that the observations could be explained
without the IMF with $\alpha=1$. 
Here we try to determine the  influence of the 
variations of the IMF on the evolution of compact binaries
and apply our results to seven regions of starformation (Shaerer et al.,
1998, hereafter SCK98).

Previously (Lipunov et al., 1996a) we used  the
``Scenario Machine'' for  calculations of  populations of
X-- ray sources after a burst of starformation
at the Galactic center. Here, as before in Popov et al.
(1997, 1998), we model a general
situation --- we make calculations for a typical starformation burst.
We show results on 
twelve types of binary sources with significant X-ray luminosity for three
values of the upper mass limit for three values of $\alpha$.  

\section{Model.}

Monte-Carlo method for statistical simulations of binary evolution
was originally proposed by Kornilov \& Lipunov (1983a,b) for massive
binaries
and developed later by Lipunov \& Postnov (1987) for low-massive binaries.
Dewey \& Cordes (1987) applied an analogous method
for analysis of radio pulsar statistics, and de Kool (1992)
investigated  by the Monte-Carlo method  the formation  of the galactic
cataclysmic variables (see the review in van den Heuvel 1994). 

Monte-Carlo simulations of binary star evolution allows one to
investigate the evolution of a large ensemble of binaries  and to
estimate the number of binaries at different
evolutionary stages. Inevitable simplifications in the
analytical description of the binary evolution that we allow in our
extensive numerical calculations, make those numbers
approximate to a factor of 2-3.  However, the inaccuracy of direct
calculations  giving the numbers of different binary types
in the Galaxy (see e.g. Iben \& Tutukov 1984, 
van den Heuvel 1994) seems to be comparable to what follows from the
simplifications in the binary evolution treatment.  

In our analysis of binary evolution, we use the ``Scenario Machine'', a
computer code, that incorporates current scenarios of binary
evolution 
and takes into account the influence of magnetic field of
compact objects on their observational appearance. A detailed description
of the computational techniques and input assumptions is summarized
elsewhere (Lipunov et al. 1996b; see also:
{\bf http://xray.sai.msu.su/\~ \, mystery/articles/review/}), 
and here we briefly list only principal parameters and initial distributions.

We trace the evolution of binary systems during the first 20 Myrs after
their formation in a starformation burst. Obviously, only stars that are
massive enough (with masses $\ge 8-10~ {\rm M}_\odot$) can evolve off
the main sequence during the time as short as this to yield compact
remnants: neutron stars (NSs) and black holes (BHs).
Therefore we consider only massive binaries, i.e. those having the mass of
the primary (more massive) component in the range of $10-120$
${\rm M}_\odot$.

The distribution in orbital separations is taken as deduced from
observations:
\begin{equation}
f(\log a) ={\rm const}\,,\qquad \max~\{10~ {\rm R}_\odot,~\hbox{Roche
Lobe}~M
(M_1)\} < \log a < 10^7~{\rm R}_\odot.
\end{equation}

We assume that a NS with a mass of $1.4~ {\rm M}_{\odot}$ is formed
as a result of the collapse of a star, whose core mass prior to collapse was
$M_*\sim (2.5-35)~{\rm M}_{\odot}$. This corresponds to an initial mass 
range $\sim (10 - 60)~{\rm M}_{\odot}$, taking into account that a massive
star can lose more than $\sim (10-20)\%$ of its initial mass during the   
evolution with a strong stellar wind.

The most massive stars are assumed to collapse into a BH once
their mass before the collapse is $M>M_{cr}=35~ {\rm M}_\odot$ (which
would correspond to an initial mass of the ZAMS star as high as $\sim
60~ {\rm M}_\odot$ since a substantial mass loss due to a strong
stellar wind occurs for the most massive stars).  The BH mass is
calculated as $M_{bh}=k_{bh}M_{cr}$, where the parameter $k_{bh}$ is
taken to be 0.7.

The mass limit for NS (the Oppenheimer-Volkoff limit) is taken to be
$M_{OV}=2.5~ {\rm M}_\odot$, which corresponds to a hard equation of
state of the NS matter.

We made calculations for several values of the coefficient $\alpha$:

\begin{equation}
   \frac{dN}{dM} \propto M^{-\alpha}
\end{equation}

We calculated $10^7$ systems in every run of the program.
Then the results were normalized to the total mass of binary stars 
in the starformation burst.
We also used different values of the upper mass limit.

We took into account that the collapse of a massive star into a NS
can be asymmetrical, so that
an additional kick velocity, $v_{kick}$, presumably randomly oriented in
space, should be imparted to the newborn compact object.
We used the velocity distribution in the form
obtained by Lyne \& Lorimer (1994) with the characteristic value 200 km/s
(twice less than in Lyne \& Lorimer (1994), see Lipunov et al. (1996c)).

\newpage

\section{Results.}

 On the figures we show the results of our calculations.
On all graphs on the X- axis we show the time after the 
starformation burst in Myrs, on the Y- axis --- number of 
the  sources of the selected type that exist at the particular moment
(not the birth rate of the sources!).

On figures 1-3 we show our calculations for X-ray sources of 12
different types for different parameters of the IMF.

\begin{itemize}

\item
Figure 1 --- $\alpha=1$, 

\item
Figure 2 --- $\alpha=1.35$,

\item
Figure 3 --- $\alpha=2.35$. 

\end{itemize}

\noindent
For upper mass limits:

\begin{itemize}

\item
$120 M_{\odot}$ -- solid lines,

\item
$60 M_{\odot}$ -- dashed lines,

\item
$40 M_{\odot}$ -- dotted lines.

\end{itemize}

 The calculated numbers were normalized for $1\cdot 10^6 
\, M_{\odot}$ in binary stars. We show on the figures 1-3
and in tables 1-9
only systems with the luminosity of compact object greater than $10^{33}\, erg/s$
(it should be mainly X-ray luminosity). 

Curves were not smoothed so all fluctuations of statistical nature are
presented. We calculated $10^7$ binary systems in every run,
and then the results were normalized.

We used the  ``flat'' mass ratio function, i.e. binary systems with
any mass ratio appear with the same probability. The results can be 
renormalized to any other form of the mass ratio function.

\section{Application of our calculations}

 We apply our results to seven regions of recent starformation.
Ages, total masses and some other characteristics were taken from SCK98
(we used total masses determined for Salpeter's IMF even for the IMFs with
different parameters, which is a simplification).
As far as for several regions ages are uncertain, we made calculations for 
two values of the age, marked in SCK98.

 Results are presented in tables 1-9 (regions NGC3125A and NGC3125B have
similar ages and total masses). We made an assumption, that binaries contain
50\% of the total mass of the starburst. Numbers were rounded off to the
nearest integer  (i.e. {\it n} sources means, that calculated number was
between {\it n}-0.5 and {\it n}+0.5).

\newpage

\section{Discussion and conclusions}

Different types of close binaries show different sensitivity to variations
of the IMF. When we replace $\alpha=2.35$ by $\alpha=1$ the numbers
of all sources increase. Systems with BHs are more sensitive to such variations. 

When one try to vary the upper mass limit, another situation appear.
In some cases (especially for $\alpha=2.35$) systems with NSs show
little differences for different values of the upper mass limit, 
while systems with BHs become significantly  less (or more) abundant
for different upper masses.
Luckily, X-ray transients, which are the most numerous systems in our
calculations,
show significant sensitivity to variations of the upper mass limit.   
But of course due to their transient nature it is difficult to
use them to detect small variations in the IMF.
If it is possible to distinguish systems with BH, it is much better to use
them to test the IMF. 

The results of our calculations can be easily used to estimate 
the number of X- ray sources for different parameters of the IMF
if the total mass of stars and age of a starburst are known 
(in (Popov et al., 1997, 1998)
analytical approximations for source numbers were given). 
And we estimate numbers of different
sources for several regions of recent starformation (tables 1-9).

In this poster we also tried to show, that, as expected, 
populations of close binaries are very sensitive
to the variations of the IMF. One must be careful,
when trying to fit the observed data for single stars
with  variations of the IMF. And, vice versa, using detailed observations of
X-ray sources, one can try to estimate parameters of the IMF, and test
results, obtained from single stars population.

\section{Acknowledgements}

 We want to thank Dr. K.A. Postnov for discussions and
G.V. Lipunova and Dr. I.E. Panchenko for technical assistance.

 This work was supported by the grants: NTP ``Astronomy'' 1.4.2.3., 
NTP ``Astronomy'' 1.4.4.1 and ``Universities of Russia'' N5559.

We are also thankful to the organizers of the conference for 
support and hospitality.

\newpage
{\Large

\begin{center}

{\bf TWELVE TYPES OF X-RAY SOURCES}

\end{center}

\noindent
{\bf BH+N2} ---  A BH with a He-core Star (Giant)
\bigskip

\noindent
{\bf NA+N1} ---  An Accreting NS with a Main Sequence Star
                (Be-transient) 
\bigskip

\noindent
{\bf BH+WR} ---  A BH with a Wolf--Rayet Star
\bigskip

\noindent
{\bf BH+N1} ---   A BH with a Main Sequence Star
\bigskip

\noindent
{\bf BH+N3G} ---  A BH with a Roche-lobe filling star, when the
binary loses angular momentum by grav. radiation
\bigskip

\noindent
{\bf NA+N3} --- An Accreting NSt with a Roche-lobe filling star
               (fast mass transfer from the more massive star)
\bigskip

\noindent
{\bf NA+WR} --- An Accreting NS with a Wolf--Rayet Star
\bigskip

\noindent
{\bf BH+N3E} --- A BH with a Roche-lobe filling star (nuclear
evolution time scale)
\bigskip

\noindent
{\bf NA+N3G} ---  An Accreting NS with a Roche-lobe filling
                  star, when the binary loses angular momentum due to 
                   gravitational radiation
\bigskip

\noindent
{\bf NA+N3M} ---  An Accreting NS with a Roche-lobe filling
                  star, when the binary loses angular 
                  momentum due to magnetic wind
\bigskip

\noindent
{\bf NA+N2} ---  An Accreting NS with a He-core Star (Giant)
\bigskip

\noindent
{\bf NA+N3E} --- An Accreting NS with a Roche-lobe filling star
(nuclear evolution time scale)

}

\newpage

{
\hbox{\hbox to 0.48\hsize{\vbox to 0.9\vsize {\epsfxsize=0.48\hsize
\epsfbox{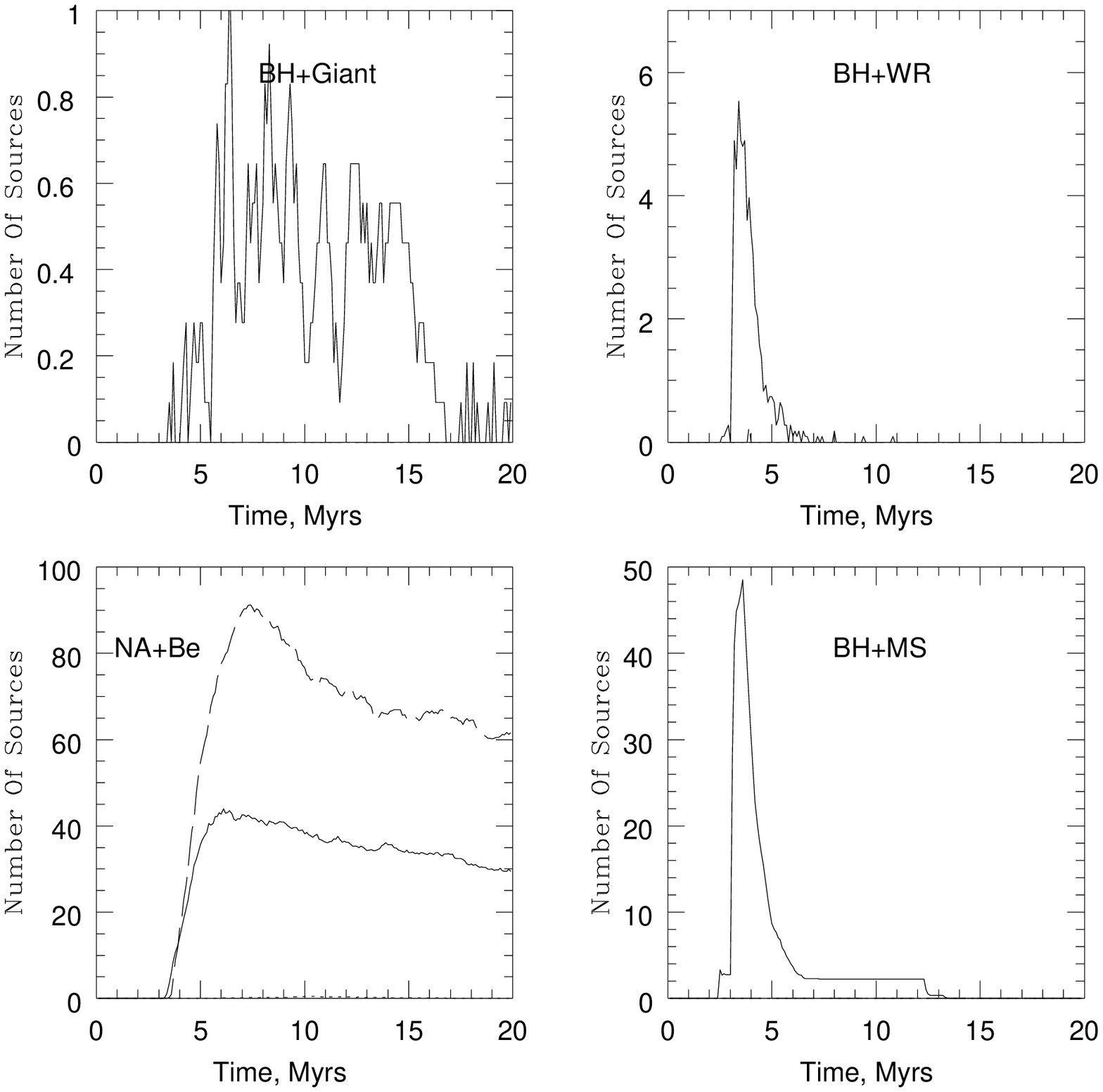} \vfill \epsfxsize=0.5\hsize \epsfbox{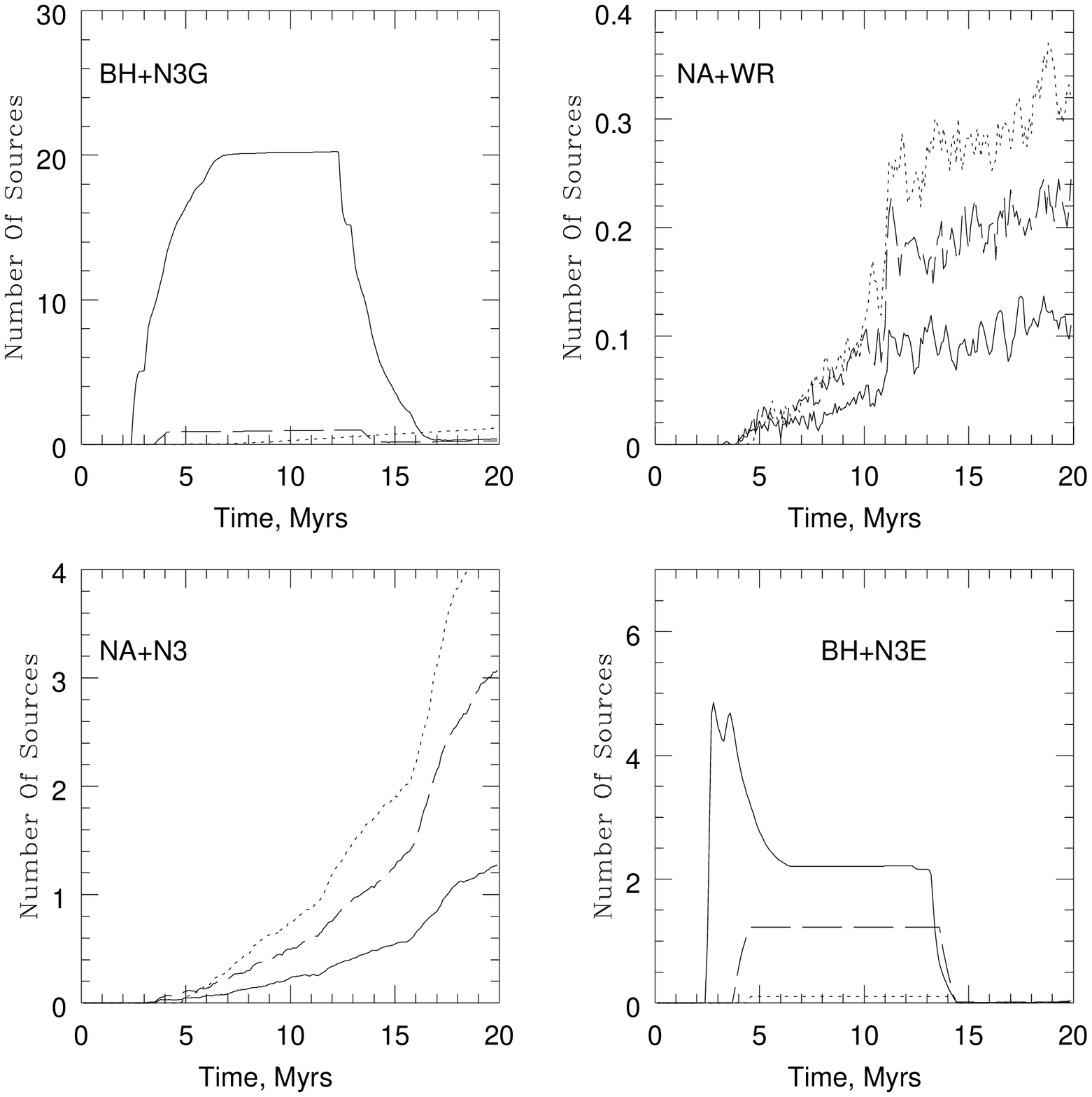}}} 
\hss \hbox to 0.48\hsize{\vbox to 0.9\vsize{\vfill \epsfxsize=0.48\hsize
\epsfbox{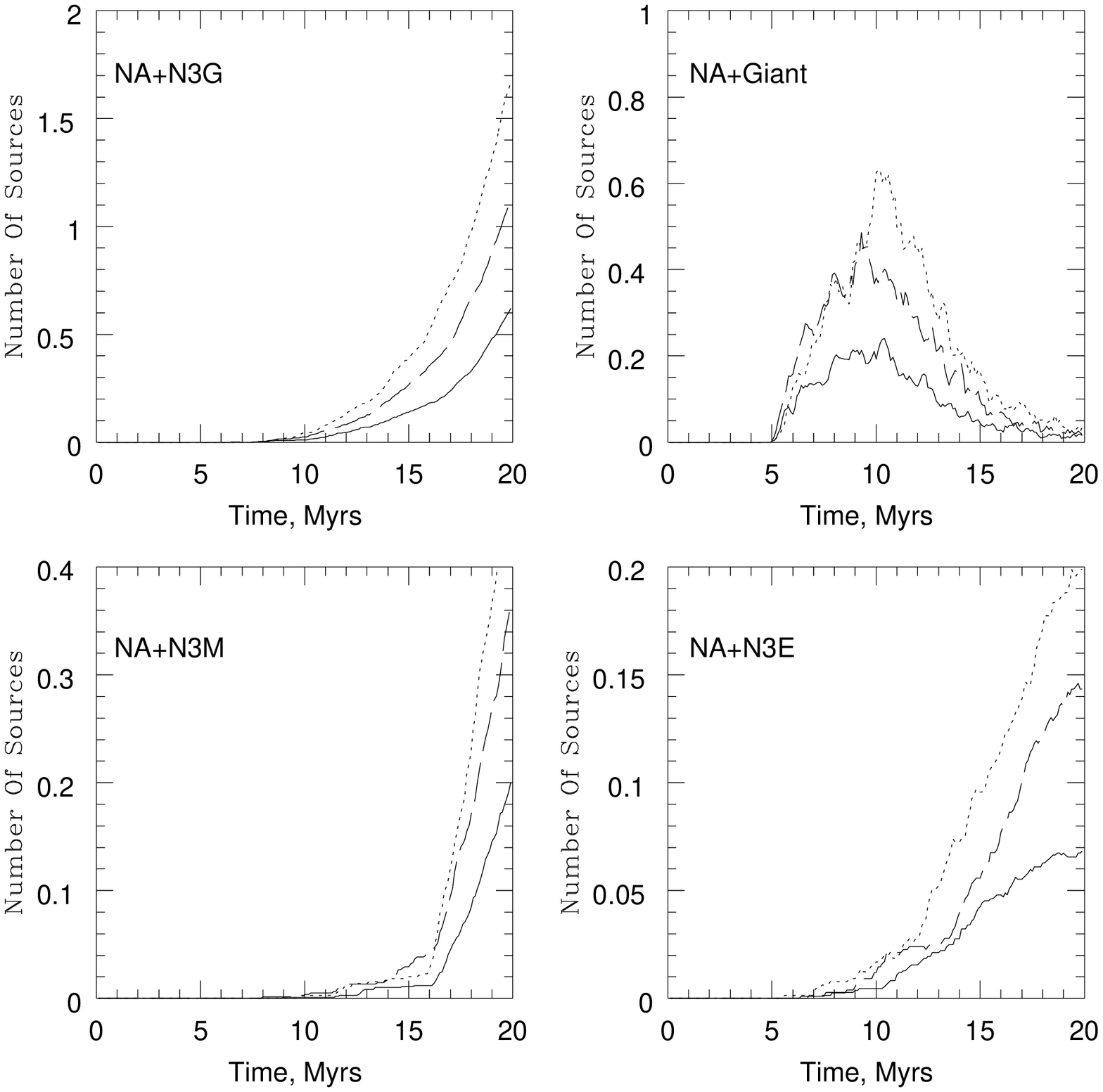} \vfill}}} 

\centerline{\bf Figure 1}
}

{
\hbox{\hbox to 0.48\hsize{\vbox to 0.9\vsize {\epsfxsize=0.48\hsize
\epsfbox{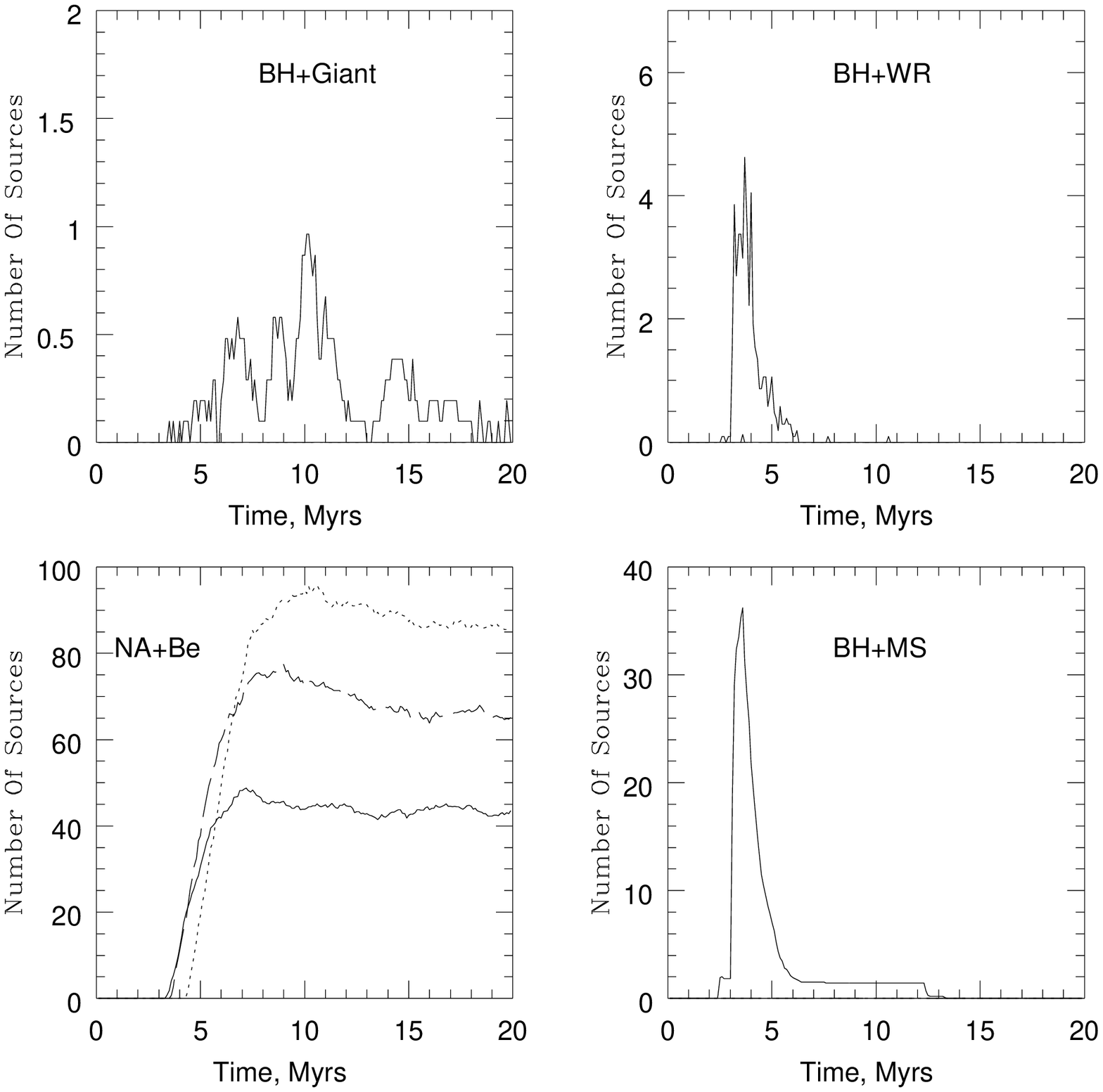} \vfill \epsfxsize=0.5\hsize \epsfbox{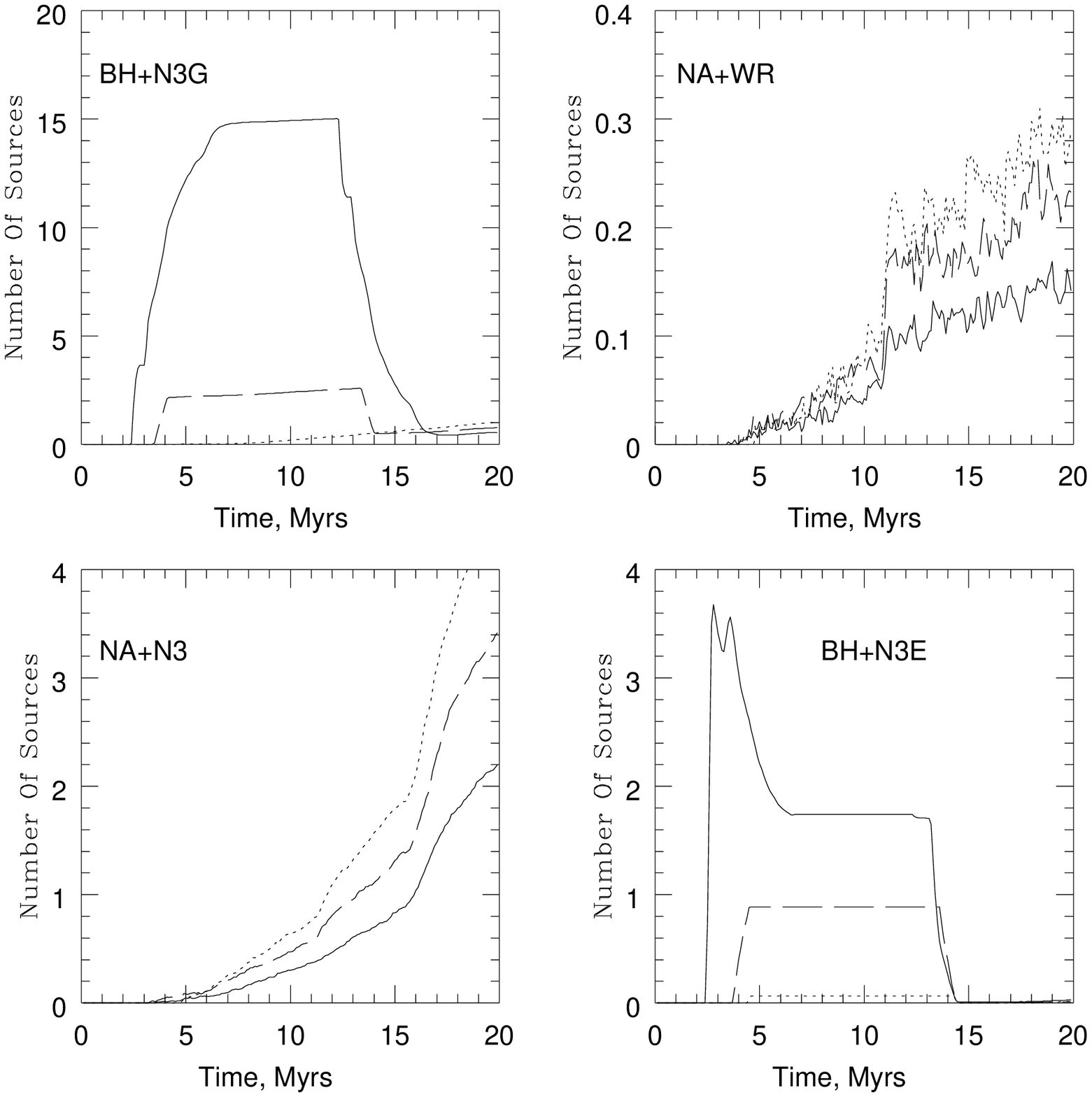}}}
\hss \hbox to 0.48\hsize{\vbox to 0.9\vsize{\vfill \epsfxsize=0.48\hsize
\epsfbox{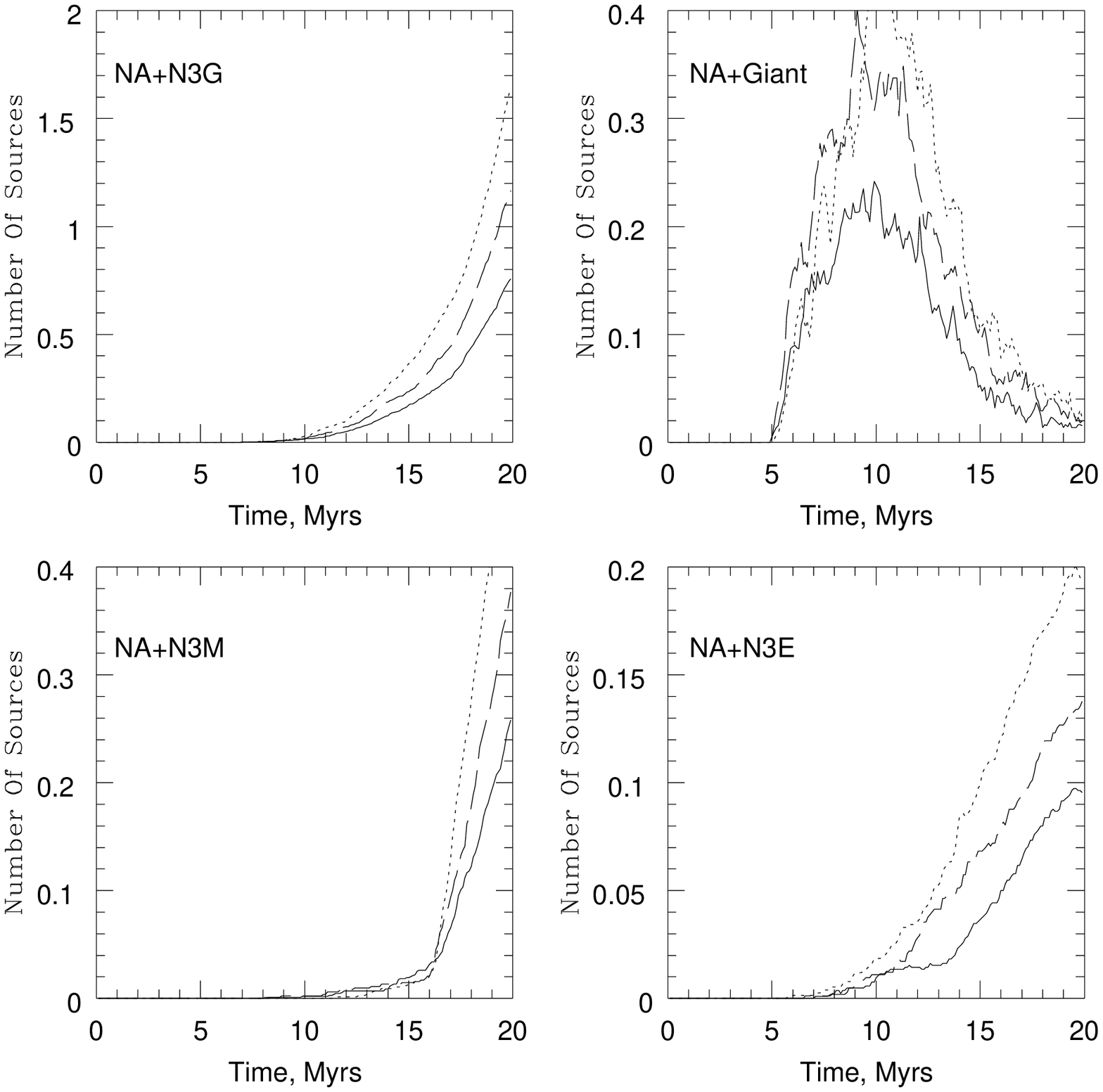} \vfill}}}

\centerline{\bf Figure 2}
}

{
\hbox{\hbox to 0.48\hsize{\vbox to 0.9\vsize {\epsfxsize=0.48\hsize
\epsfbox{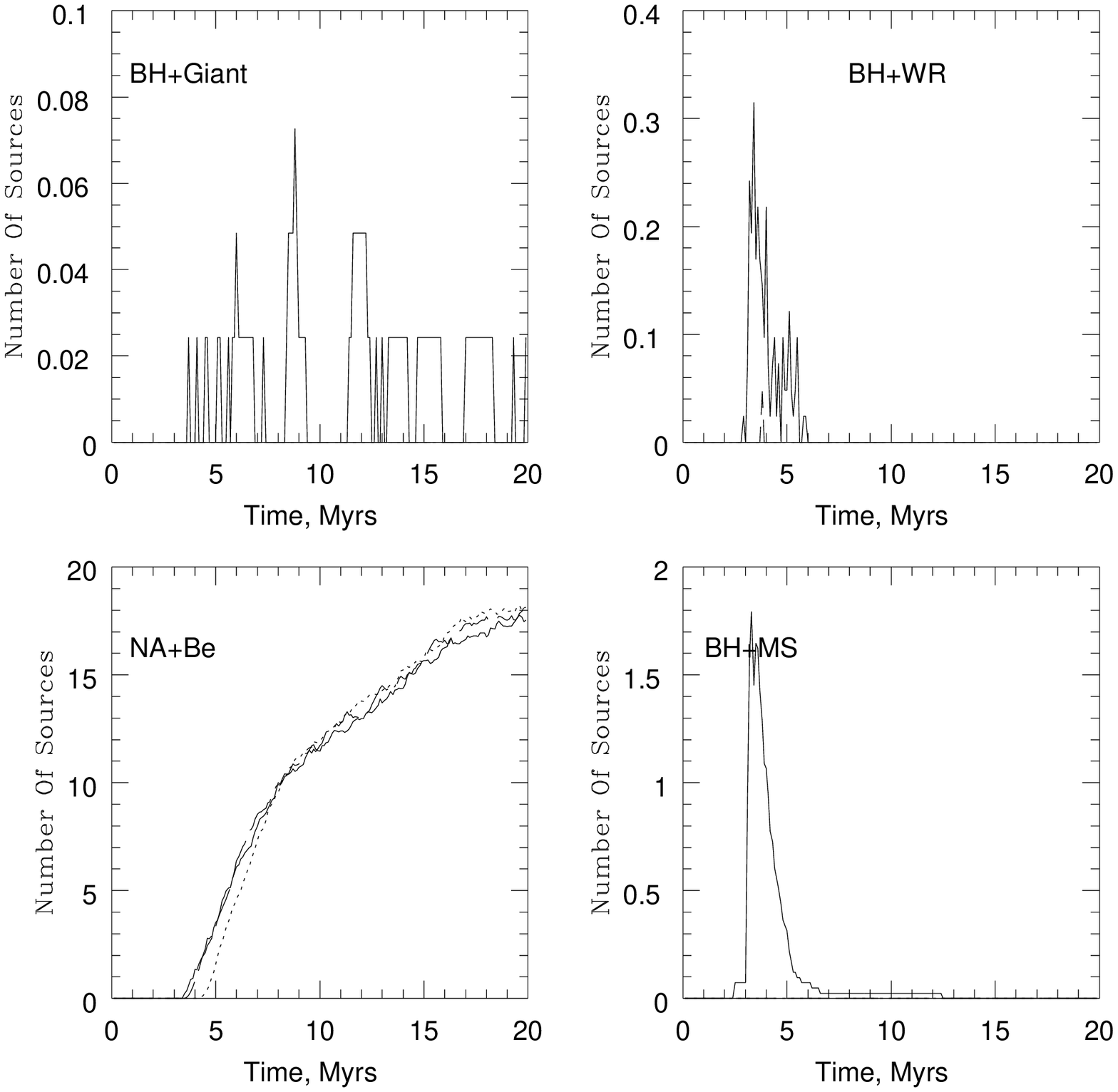} \vfill \epsfxsize=0.5\hsize \epsfbox{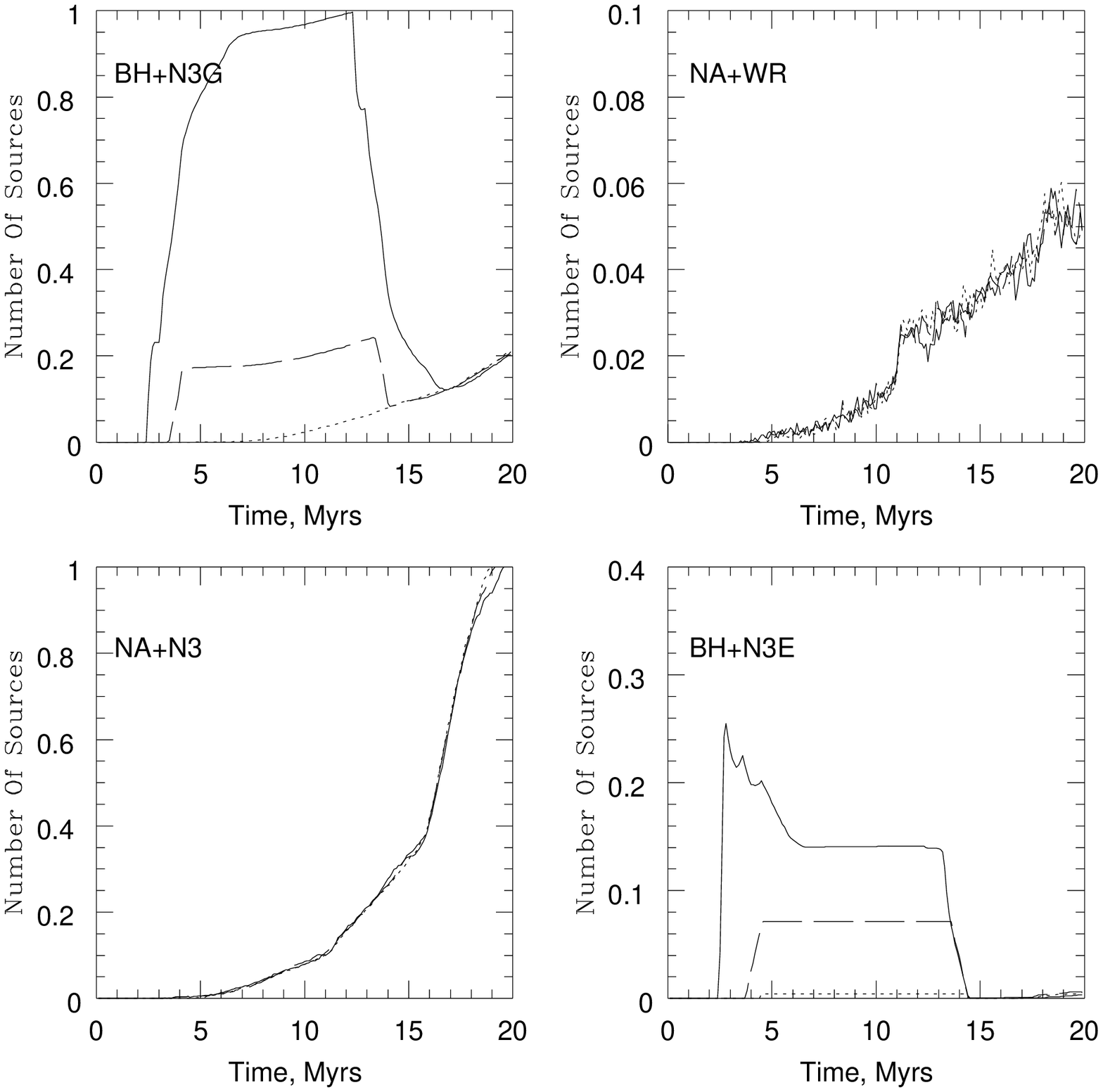}}}
\hss \hbox to 0.48\hsize{\vbox to 0.9\vsize{\vfill \epsfxsize=0.48\hsize
\epsfbox{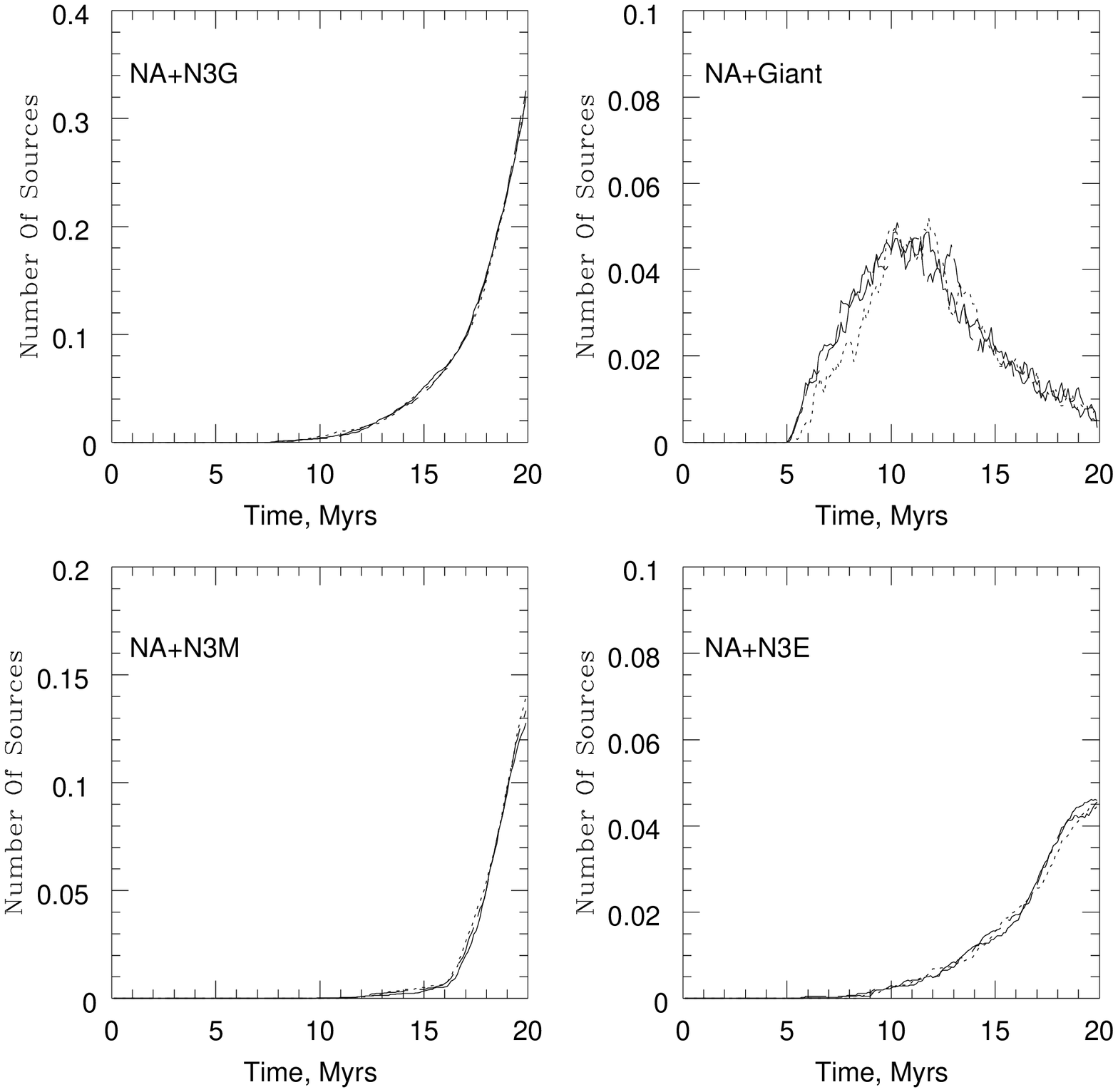} \vfill}}}

\centerline{\bf Figure 3}
}

\newpage

\begin{table*}[h]
\caption[]{He 2-10;  age 5.5 Myrs; total mass $10^{6.8} M_{\odot}$}
\begin{tabular}{|l||c|c|c|c|c|c|c|c|c|}
\hline
 Slope&2.35&2.35&2.35& 1.35&1.35&1.35&  1.01&1.01&1.01\\
\hline     

Up.mas.&120&  60&  40&   120&  60&  40&   120&  60& 40\\

%\hline
\hline
bh+n1 & 0 &  0 &  0 &   16 &  0 &  0 &   28 &  0 &  0\\  

bh+n2 & 0 &  0 &  0 &    0 &  0 &  0 &    0 &  0 &  0\\

bh+n3e& 1 &  0 &  0 &    9 &  4 &  0 &   12 &  6 &  1\\

bh+n3g& 4 &  1 &  0 &   62 & 10 &  0 &   84 &  4 &  0\\

bh+wr & 0 &  0 &  0 &    1 &  0 &  0 &    3 &  0 &  0\\

na+n1 &24 & 22 & 15 &  187 &241 &165 &  190 &321 &221\\

na+n3 & 0 &  0 &  0 &    0 &  0 &  0 &    0 &  1 &  0\\

na+wr & 0 &  0 &  0 &    0 &  0 &  0 &    0 &  0 &  0\\

na+n3m& 0 &  0 &  0 &    0 &  0 &  0 &    0 &  0 &  0\\

na+n3e& 0 &  0 &  0 &    0 &  0 &  0 &    0 &  0 &  0\\

na+n3g& 0 &  0 &  0 &    0 &  0 &  0 &    0 &  0 &  0\\

na+n2 & 0 &  0 &  0 &    0 &  0 &  0 &    0 &  0 &  0\\

\hline
\end{tabular}
\end{table*}

\begin{table*}[h]
\caption[]{He 2-10;  age 6.0 Myrs; total mass $10^{6.8} M_{\odot}$}
\begin{tabular}{|l||c|c|c|c|c|c|c|c|c|}
\hline
 Slope&2.35&2.35&2.35& 1.35&1.35&1.35&  1.01&1.01&1.01\\
\hline     

Up.mas.&120&  60&  40&   120&  60&  40&   120&  60& 40\\

%\hline
\hline
bh+n1 &  0&  0 &  0 &   9  & 0  & 0  & 17&   0&   0\\

bh+n2 &  0&  0 &  0 &   1  & 0  & 0  &  2&   0&   0\\

bh+n3e&  1&  0 &  0 &   9  & 4  & 0  & 11&   6&   1\\

bh+n3g&  4&  1 &  0 &  65  &11  & 0  & 88&   4&   0\\

bh+wr &  0&  0 &  0 &   0  & 0  & 0  &  0&   0&   0\\

na+n1 & 29& 30 & 22 & 198  &283 &233 &202& 367& 322\\

na+n3 &  0&  0 &  0 &   0  & 1  & 1  &  0&   1&   1\\

na+wr &  0&  0 &  0 &   0  & 0  & 0  &  0&   0&   0\\

na+n3m&  0&  0 &  0 &   0  & 0  & 0  &  0&   0&   0\\

na+n3e&  0&  0 &  0 &   0  & 0  & 0  &  0&   0&   0\\

na+n3g&  0&  0 &  0 &   0  & 0  & 0  &  0&   0&   0\\

na+n2 &  0&  0 &  0 &   0  & 1  & 0  &  0&   1&   1\\

\hline
\end{tabular}
\end{table*}

\clearpage

\begin{table*}[h]
\caption[]{NGC3125A, B;  age 4.5 Myrs; total mass $10^{6.1} M_{\odot}$}
\begin{tabular}{|l||c|c|c|c|c|c|c|c|c|}
\hline
 Slope&2.35&2.35&2.35& 1.35&1.35&1.35&  1.01&1.01&1.01\\
\hline     

Up.mas.&120&  60&  40&   120&  60&  40&   120&  60& 40\\

\hline
%\hline
bh+n1 & 1 &  0 &  0 &   11 &  0 &  0 &   16 &  0 &  0\\

bh+n2 & 0 &  0 &  0 &    0 &  0 &  0 &    0 &  0 &  0\\

bh+n3e& 0 &  0 &  0 &    2 &  1 &  0 &    3 &  1 &  0\\

bh+n3g& 1 &  0 &  0 &   11 &  2 &  0 &   14 &  1 &  0\\

bh+wr & 0 &  0 &  0 &    1 &  0 &  0 &    1 &  0 &  0\\

na+n1 & 2 &  2 &  0 &   21 & 24 &  5 &   24 & 33 &  6\\

na+n3 & 0 &  0 &  0 &    0 &  0 &  0 &    0 &  0 &  0\\

na+wr & 0 &  0 &  0 &    0 &  0 &  0 &    0 &  0 &  0\\

na+n3m& 0 &  0 &  0 &    0 &  0 &  0 &    0 &  0 &  0\\

na+n3e& 0 &  0 &  0 &    0 &  0 &  0 &    0 &  0 &  0\\

na+n3g& 0 &  0 &  0 &    0 &  0 &  0 &    0 &  0 &  0\\

na+n2 & 0 &  0 &  0 &    0 &  0 &  0 &    0 &  0 &  0\\
\hline

\end{tabular}
\end{table*}

\begin{table*}[h]
\caption[]{NGC3125A, B;  age 5.0 Myrs; total mass $10^{6.1} M_{\odot}$}
\begin{tabular}{|l||c|c|c|c|c|c|c|c|c|}
\hline
 Slope&2.35&2.35&2.35& 1.35&1.35&1.35&  1.01&1.01&1.01\\
\hline

Up.mas.&120&  60&  40&   120&  60&  40&   120&  60& 40\\

\hline     
%\hline
bh+n1 & 0 &  0 &  0 &    7 &  0 &  0 &    8 &  0 &  0\\

bh+n2 & 0 &  0 &  0 &    0 &  0 &  0 &    0 &  0 &  0\\

bh+n3e& 0 &  0 &  0 &    2 &  1 &  0 &    3 &  1 &  0\\

bh+n3g& 1 &  0 &  0 &   12 &  2 &  0 &   16 &  1 &  0\\

bh+wr & 0 &  0 &  0 &    1 &  0 &  0 &    1 &  0 &  0\\

na+n1 & 3 &  3 &  1 &   29 & 36 & 18 &   34 & 52 & 25\\

na+n3 & 0 &  0 &  0 &    0 &  0 &  0 &    0 &  0 &  0\\

na+wr & 0 &  0 &  0 &    0 &  0 &  0 &    0 &  0 &  0\\

na+n3m& 0 &  0 &  0 &    0 &  0 &  0 &    0 &  0 &  0\\

na+n3e& 0 &  0 &  0 &    0 &  0 &  0 &    0 &  0 &  0\\

na+n3g& 0 &  0 &  0 &    0 &  0 &  0 &    0 &  0 &  0\\

na+n2 & 0 &  0 &  0 &    0 &  0 &  0 &    0 &  0 &  0\\

\hline
\end{tabular}
\end{table*}

\clearpage

\begin{table*}[h]
\caption[]{NGC5253A; age 3.0 Myrs; total mass $10^{6.6} M_{\odot}$}
\begin{tabular}{|l||c|c|c|c|c|c|c|c|c|}
\hline
 Slope&2.35&2.35&2.35& 1.35&1.35&1.35&  1.01&1.01&1.01\\
\hline

Up.mas.&120&  60&  40&   120&  60&  40&   120&  60& 40\\

\hline     
%\hline
bh+n1 & 0 &  0 &  0 &    5 &  0 &  0 &    8 &  0 &  0\\

bh+n2 & 0 &  0 &  0 &    0 &  0 &  0 &    0 &  0 &  0\\

bh+n3e& 1 &  0 &  0 &   10 &  0 &  0 &   13 &  0 &  0\\

bh+n3g& 1 &  0 &  0 &   11 &  0 &  0 &   15 &  0 &  0\\

bh+wr & 0 &  0 &  0 &    0 &  0 &  0 &    0 &  0 &  0\\

na+n1 & 0 &  0 &  0 &    0 &  0 &  0 &    0 &  0 &  0\\

na+n3 & 0 &  0 &  0 &    0 &  0 &  0 &    0 &  0 &  0\\

na+wr & 0 &  0 &  0 &    0 &  0 &  0 &    0 &  0 &  0\\

na+n3m& 0 &  0 &  0 &    0 &  0 &  0 &    0 &  0 &  0\\

na+n3e& 0 &  0 &  0 &    0 &  0 &  0 &    0 &  0 &  0\\

na+n3g& 0 &  0 &  0 &    0 &  0 &  0 &    0 &  0 &  0\\

na+n2 & 0 &  0 &  0 &    0 &  0 &  0 &    0 &  0 &  0\\
\hline

\end{tabular}
\end{table*}

\begin{table*}[h]
\caption[]{NGC5253B; age 5.0 Myrs; total mass $10^{6.6} M_{\odot}$}
\begin{tabular}{|l||c|c|c|c|c|c|c|c|c|}
\hline
 Slope&2.35&2.35&2.35& 1.35&1.35&1.35&  1.01&1.01&1.01\\
\hline     

Up.mas.&120&  60&  40&   120&  60&  40&   120&  60& 40\\

%\hline
\hline
bh+n1 & 1 &  0 &  0 &   21&   0 &  0 &   26&   0 &  0\\

bh+n2 & 0 &  0 &  0 &    1&   0 &  0 &    1&   0 &  0\\

bh+n3e& 1 &  0 &  0 &    7&   3 &  0 &    8&   4 &  0\\

bh+n3g& 2 &  1 &  0 &   36&   7 &  0 &   49&   3 &  0\\

bh+wr & 0 &  0 &  0 &    3&   0 &  0 &    2&   0 &  0\\

na+n1 &11 & 10 &  5 &   92& 112 & 58 &  106& 163 & 80\\

na+n3 & 0 &  0 &  0 &    0&   0 &  0 &    0&   0 &  0\\

na+wr & 0 &  0 &  0 &    0&   0 &  0 &    0&   0 &  0\\

na+n3m& 0 &  0 &  0 &    0&   0 &  0 &    0&   0 &  0\\

na+n3e& 0 &  0 &  0 &    0&   0 &  0 &    0&   0 &  0\\

na+n3g& 0 &  0 &  0 &    0&   0 &  0 &    0&   0 &  0\\

na+n2 & 0 &  0 &  0 &    0&   0 &  0 &    0&   0 &  0\\
\hline

\end{tabular}
\end{table*}

\clearpage

\begin{table*}[h]
\caption[]{Tol 89; age 4.5 Myrs; total mass $10^{5.7} M_{\odot}$}
\begin{tabular}{|l||c|c|c|c|c|c|c|c|c|}
\hline
 Slope&2.35&2.35&2.35& 1.35&1.35&1.35&  1.01&1.01&1.01\\
\hline     

Up.mas.&120&  60&  40&   120&  60&  40&   120&  60& 40\\

\hline
%\hline
bh+n1 & 0 &  0 &  0 &    4 &  0 &  0 &    6 &  0 &  0\\

bh+n2 & 0 &  0 &  0 &    0 &  0 &  0 &    0 &  0 &  0\\

bh+n3e& 0 &  0 &  0 &    1 &  0 &  0 &    1 &  0 &  0\\

bh+n3g& 0 &  0 &  0 &    4 &  1 &  0 &    6 &  0 &  0\\

bh+wr & 0 &  0 &  0 &    0 &  0 &  0 &    1 &  0 &  0\\

na+n1 & 1 &  1 &  0 &    9 &  9 &  2 &   10 & 13 &  2\\

na+n3 & 0 &  0 &  0 &    0 &  0 &  0 &    0 &  0 &  0\\

na+wr & 0 &  0 &  0 &    0 &  0 &  0 &    0 &  0 &  0\\

na+n3m& 0 &  0 &  0 &    0 &  0 &  0 &    0 &  0 &  0\\

na+n3e& 0 &  0 &  0 &    0 &  0 &  0 &    0 &  0 &  0\\

na+n3g& 0 &  0 &  0 &    0 &  0 &  0 &    0 &  0 &  0\\

na+n2 & 0 &  0 &  0 &    0 &  0 &  0 &    0 &  0 &  0\\

\hline
\end{tabular}
\end{table*}

\begin{table*}[h]
\caption[]{Tol 89;  age 5.0 Myrs; total mass $10^{5.7} M_{\odot}$}
\begin{tabular}{|l||c|c|c|c|c|c|c|c|c|}
\hline
 Slope&2.35&2.35&2.35& 1.35&1.35&1.35&  1.01&1.01&1.01\\
\hline     

Up.mas.&120&  60&  40&   120&  60&  40&   120&  60& 40\\

\hline
%\hline
bh+n1 & 0 &  0 &  0 &    3 &  0 &  0 &    3 &  0 &  0\\

bh+n2 & 0 &  0 &  0 &    0 &  0 &  0 &    0 &  0 &  0\\

bh+n3e& 0 &  0 &  0 &    1 &  0 &  0 &    1 &  0 &  0\\

bh+n3g& 0 &  0 &  0 &    5 &  1 &  0 &    6 &  0 &  0\\

bh+wr & 0 &  0 &  0 &    0 &  0 &  0 &    0 &  0 &  0\\

na+n1 & 1 &  1 &  1 &   12 & 14 &  7 &   13 & 21 & 10\\

na+n3 & 0 &  0 &  0 &    0 &  0 &  0 &    0 &  0 &  0\\

na+wr & 0 &  0 &  0 &    0 &  0 &  0 &    0 &  0 &  0\\

na+n3m& 0 &  0 &  0 &    0 &  0 &  0 &    0 &  0 &  0\\

na+n3e& 0 &  0 &  0 &    0 &  0 &  0 &    0 &  0 &  0\\

na+n3g& 0 &  0 &  0 &    0 &  0 &  0 &    0 &  0 &  0\\

na+n2 & 0 &  0 &  0 &    0 &  0 &  0 &    0 &  0 &  0\\

\hline
\end{tabular}
\end{table*}

\clearpage

\begin{table*}[h]
\caption[]{NGC3049; age 5.5 Myrs; total mass $10^{6.4} M_{\odot}$}
\begin{tabular}{|l||c|c|c|c|c|c|c|c|c|}
\hline
 Slope&2.35&2.35&2.35& 1.35&1.35&1.35&  1.01&1.01&1.01\\
\hline     

Up.mas.&120&  60&  40&   120&  60&  40&   120&  60& 40\\

\hline
%\hline
bh+n1 & 0 &  0 &  0 &    7 &  0 &  0 &   11&   0 &  0\\

bh+n2 & 0 &  0 &  0 &    0 &  0 &  0 &    0&   0 &  0\\

bh+n3e& 0 &  0 &  0 &    4 &  2 &  0 &    5&   2 &  0\\

bh+n3g& 2 &  0 &  0 &   24 &  4 &  0 &   33&   2 &  0\\

bh+wr & 0 &  0 &  0 &    1 &  0 &  0 &    1&   0 &  0\\

na+n1 & 9 &  9 &  6 &   74 & 96 & 66 &   76& 128 & 88\\

na+n3 & 0 &  0 &  0 &    0 &  0 &  0 &    0&   0 &  0\\

na+wr & 0 &  0 &  0 &    0 &  0 &  0 &    0&   0 &  0\\

na+n3m& 0 &  0 &  0 &    0 &  0 &  0 &    0&   0 &  0\\

na+n3e& 0 &  0 &  0 &    0 &  0 &  0 &    0&   0 &  0\\

na+n3g& 0 &  0 &  0 &    0 &  0 &  0 &    0&   0 &  0\\

na+n2 & 0 &  0 &  0 &    0 &  0 &  0 &    0&   0 &  0\\

\hline
\end{tabular}
\end{table*}

\clearpage

\end{document}